\documentclass[reprint,prb,aps,amsmath,amssymb,superscriptaddress]{revtex4-2}

\usepackage[T1]{fontenc}
\usepackage[utf8]{inputenc}
\usepackage[english]{babel}
\usepackage{bm}
\usepackage{graphicx}
\usepackage{units}
\usepackage{orcidlink}
\usepackage{hyperref}
\hypersetup{
  colorlinks=true,
  citecolor=blue,
  urlcolor=blue,
  linkcolor=red,
  breaklinks=true
}
\begin{document}

\title{Mode-Resolved Multiband Ballistic Transport and Conductance Thresholds in Bilayer Graphene Junctions}
\author{Dan-Na Liu
\orcidlink{0009-0009-9351-2104}
}
\affiliation{School of Physics, Xidian University, Xi$'$an, Shaanxi 710071, China}
\affiliation{IMDEA Nanoscience, C/ Faraday 9, 28049 Madrid, Spain}
\author{Jun Zheng
\orcidlink{0000-0001-8426-473X}
}
\affiliation{College of Physics and Technology, Bohai University, Jinzhou, Liaoning 121013, China}
\author{Pierre A. Pantale\'on
\orcidlink{0000-0003-1709-7868}
}
\email{pierre.pantaleon@imdea.org}
\affiliation{IMDEA Nanoscience, C/ Faraday 9, 28049 Madrid, Spain} 

\begin{abstract}
We study ballistic transport in bilayer graphene junctions and show how electrostatic gating, interlayer bias, and homogeneous strain provide complementary control over electron transmission. In the absence of strain, transport is governed by symmetry constraints that suppress transmission at specific incidence angles despite the availability of states. An interlayer bias lifts this suppression through mode mixing and opens a tunable transport gap. Within a full four-band description, we identify a distinct conductance threshold that marks the onset of propagation of the upper band inside the barrier. This produces a clear change in the slope of the conductance and serves as an experimentally accessible transport fingerprint of the multiband structure and interlayer coupling. Homogeneous in-plane strain acts as a geometric control mechanism. By reshaping the band structure in momentum space, it redistributes the angular transmission window and suppresses conductance without introducing disorder. Importantly, strain preserves the underlying symmetry-based decoupling responsible for transmission suppression while shifting its condition away from normal incidence. These results provide a unified framework for interpreting angle-resolved transport in bilayer graphene and establish multiband ballistic transport as a practical probe of band-structure geometry.
\end{abstract}
\maketitle
\section{Introduction}

Bilayer graphene (BG) provides a natural platform for studying ballistic transport in systems where multiple electronic modes coexist at low energies. Unlike monolayer graphene, which hosts massless Dirac fermions, AB-stacked BG exhibits a multiband spectrum with massive chiral quasiparticles and an electrically tunable band gap~\cite{mccann2006landau,castro2007biased,Varlet2015Band}. As a result, electron propagation in BG junctions generally involves several longitudinal solutions at a given energy, including both propagating and evanescent components, leading to transport behavior without direct analogue in either monolayer graphene or conventional semiconductor heterostructures~\cite{Gu2011Chirality,Yamamoto1989Transmission,Wu1991Quantum,Ulloa1990Ballistic}.

This multichannel structure has clear experimental manifestations. Electrostatic tunnel junctions in BG display resonant tunneling and negative differential resistance, emphasizing the role of interband coupling and evanescent modes~\cite{Fallahazad2014Gate,Burg2018Strongly,Gayduchenko2021Tunnel}. In lateral geometries, gate-defined cavities support ballistic Fabry-Pérot interference~\cite{Varlet2014Fabry}, while angularly sensitive transport measurements reveal that transmission depends strongly on the geometry of the underlying band structure rather than solely on barrier height or carrier density~\cite{Elahi2024Direct}. These observations establish BG junctions as systems in which ballistic transport is governed by the availability and coupling of distinct electronic modes.

From a theoretical perspective, electrostatically defined barriers provide a controlled setting for analyzing interference, chirality, and symmetry constraints in graphene-based systems~\cite{Beenaker2008Colloquium}. In BG, a full four-band description is essential: at fixed energy, two longitudinal solutions generally coexist in the leads, and transport depends on how these incident propagating modes couple to the set of propagating and evanescent solutions supported inside the electrostatic barrier~\cite{Nilsson2007Transmission,van2013four,Gu2011Chirality,Huang2025Evanescent,Liu2026Mode}. Symmetry plays a central role in this coupling. At normal incidence, specific internal modes can become symmetry-decoupled from incoming propagating states, leading to a strong suppression of transmission even though internal solutions exist at the same energy~\cite{Gu2011Chirality,Liu2026Mode}. This symmetry-imposed decoupling underlies the cloaking phenomenon in BG transport. While related suppression effects have often been discussed under the label of anti-Klein tunneling~\cite{Katsnelson2006Chiral,Varlet2014Fabry,AgrawalGarg2012Reversal,Chen2009Design,Betancur2025Topical}, a channel-resolved four-band framework is required to capture the general mode-dependent transport behavior of BG junctions.

Transport through single and multiple electrostatic barriers in BG has therefore been widely studied, revealing tunneling suppression, resonant transmission, and interference effects~\cite{Snyman2007Ballistic,Nilsson2007Transmission,van2013four,barbier2009bilayer,Lee2016Evidence,Chen2009Design}. In multibarrier geometries, Fabry-Pérot-like oscillations arise from phase-coherent propagation between successive barriers~\cite{Ulloa1990Ballistic,Wu1991Quantum}. At the same time, the existence of internal propagating solutions inside individual barriers points to additional transport regimes~\cite{Liu2026Mode} absent in reduced two-band descriptions~\cite{Gu2011Chirality,Park2011Pi}. Because most experiments probe conductance integrated over all incident angles and channels, however, the microscopic role of individual propagating and evanescent modes often remains implicit.

An additional and experimentally relevant control parameter in this context is mechanical strain. In graphene, homogeneous strain reshapes the low-energy electronic structure by modifying lattice geometry~\cite{wang2019situ,peng2020strain}, and can generate pseudomagnetic fields~\cite{vozmediano2010gauge,zhou2023imaging}, scalar potentials~\cite{He2015Tuning}, and valley-dependent phenomena~\cite{hsu2020nanoscale}. Such strain fields can be realized experimentally through substrate engineering or mechanical actuation~\cite{shioya2014straining,klimov2012electromechanical,Bunch2008Impermeable}. In AB-stacked BG, interlayer coupling amplifies the impact of lattice deformations, with homogeneous in-plane strain shifting and distorting the quadratic band-touching points and inducing anisotropy in momentum space~\cite{Georgoulea2022Strain}. While ballistic transport in strained monolayer graphene has been extensively investigated~\cite{pellegrino2012resonant,wang2023valley,lu2016valley,munoz2017analytic,sattari2016spin,sattari2020tunneling,wang2014graphene}, its consequences for mode-resolved transport in strained BG junctions remain far less explored.

In this work, we study ballistic electron transport in AB-stacked bilayer graphene junctions within a full four-band, mode-resolved framework. Using a low-energy continuum description and a strain-extended transfer-matrix formalism, we analyze transport across normal-modulated-normal junctions by explicitly resolving propagating and evanescent modes. For purely electrostatic barriers, we identify distinct transport regimes governed by mode availability and symmetry-imposed decoupling, leading to Fabry-Pérot resonances and suppression of transmission at normal incidence despite the presence of internal solutions. We further show that the multiband structure produces a characteristic conductance threshold associated with the onset of propagation of the internal high-energy branch, resulting in a clear change in conductance slope that provides a direct transport signature of the interlayer coupling. An interlayer bias lifts symmetry protection through mode mixing, opens a transport gap, and shifts this threshold. Finally, we demonstrate that homogeneous in-plane strain acts as a geometric deformation of the mode structure, reshaping isoenergetic contours, breaking angular symmetry, and displacing the cloaking condition away from normal incidence without destroying it.

The paper is organized as follows. The first section introduces the four-band model and the transfer-matrix formalism. The next section presents the mode-resolved transport results for electrostatic barriers, interlayer bias, and strain. This is followed by a momentum-space interpretation of the transmission based on isoenergetic contours. The subsequent section discusses the conductance obtained from the mode-resolved transmission, and the final section summarizes the main conclusions.

\begin{figure}
  \centering
  \includegraphics[width= 1 \columnwidth]{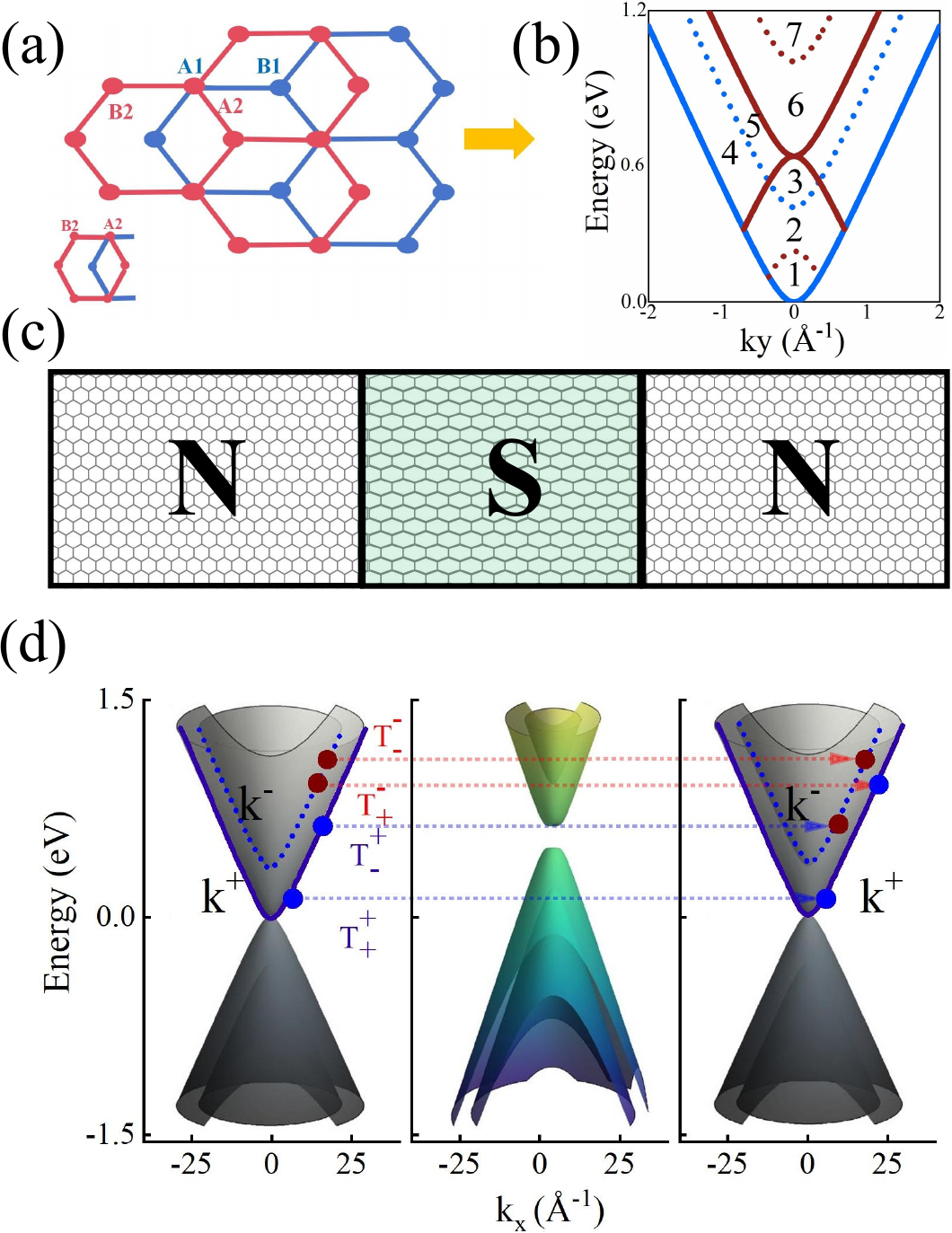}
  \caption{(a) The deformed AB-stacked BG lattice, and inset is the graphene lattice without strain. (b) In the presence of an electrostatic potential term, $V_0=0.6$ eV (red curve), the distribution of different propagating modes in the $k_{y}-E$ plane can be divided into several regions. The blue lines denote the boundaries between propagating and evanescent waves in region N, while the red lines denote the corresponding boundaries in region S. The different transport regimes are indicated from 1 to 7. (c) Illustration of a strain-modulated BG junction, where $N$ and $S$ denote the non-modulated region and the modulated region, respectively. (d) A schematic diagram of the transmission and reflection of the four band model in N-S-N BG. Here, $T^{s}_{s'}$ denotes the transmission or reflection for the propagation from mode $s$ to mode $s'$, respectively.}
  \label{fig:1}
\end{figure}\par

\begin{figure*}
  \centering 
  \includegraphics[scale=0.3]{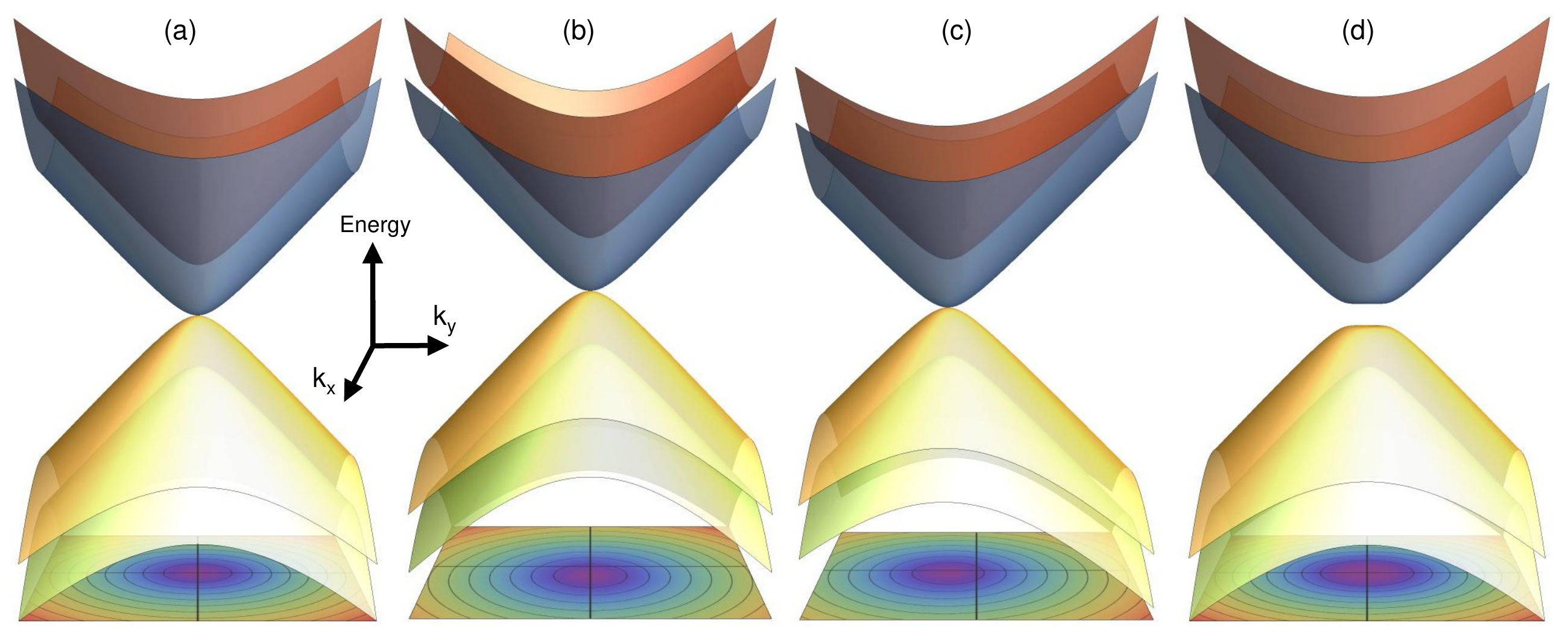}
  \caption{Electronic structure of bilayer graphene for (a) the unstrained case, (b) $\epsilon=2\%$ with $\theta=0$, and (c) $\epsilon=2\%$ with $\theta=0.2\pi$. Panel in (d) is the unstrained case with a perpendicular electric field. The bottom of each panel shows the contour plot of the lower middle band. The purple region marks the position of the quadratic band touching between the two middle bands. The two black lines indicate the trajectories corresponding to $k_{x}=0$ and $k_{y}=0$, respectively. In the absence of strain, the quadratic band touching coincides with the intersection of these two lines as shown in panel in (a). In the presence of strain, the quadratic touching is shifted.}
  \label{fig:Bandas}
\end{figure*}

\section{Model and Theoretical Framework}
\label{sec: Theory}

\subsection{Hamiltonian of BG Without Strain}

We model our device as an extended AB-stacked BG sheet divided into three regions along the transport direction $x$: an unmodulated left region ($x<0$), a central modulated region ($0<x<L$), and an unmodulated right region ($x>L$), as shown in Fig.~\ref{fig:1}(c). For convenience, these regions are denoted as N, S, and N, respectively. In the central region, we include the combined effects of a uniform electrostatic potential $V_0$~\cite{Pereira2010Klein}, an interlayer bias $V_1$~\cite{Nandkishore2011Common,Oostinga2007Gate}, and, where specified later, a homogeneous in-plane strain of magnitude $\epsilon$ applied along direction $\theta$~\cite{pellegrino2011transport}.

BG consists of two hexagonal graphene monolayers with intralayer carbon-carbon bond length $a=0.142$~nm. Each layer contains two sublattices, labeled $A_1$ and $B_1$ in the top layer and $A_2$ and $B_2$ in the bottom layer. In the AB-stacked configuration, the $A_2$ atoms lie directly above the $B_1$ atoms, as illustrated in Fig.~\ref{fig:1}(a). The dominant intralayer hopping amplitude is $\gamma_0 \approx 3$ eV, while the leading interlayer coupling between $A_2$ and $B_1$ sites is $\gamma_1 = 0.4$ eV. Additional interlayer hopping parameters $\gamma_3$ and $\gamma_4$ give rise to trigonal warping effects that become relevant only at very low energies ($E < 10$ meV). Since we focus on transport at higher energies ($E > 50$ meV), these terms are neglected throughout~\cite{mccann2006landau,Mayorov2011Interaction,Kleptsyn2015Chiral}. Within this approximation, the low-energy effective Hamiltonian near the $K$ point is given by
\begin{equation}
H =
\begin{pmatrix}
V_0 + V_1 & \hbar v_f \pi & \gamma_1 & 0 \\
\hbar v_f \pi^\dagger & V_0 + V_1 & 0 & 0 \\
\gamma_1 & 0 & V_0 - V_1 & \hbar v_f \pi^\dagger \\
0 & 0 & \hbar v_f \pi & V_0 - V_1
\label{eq: MainHamiltonian}
\end{pmatrix},
\end{equation}
written in the basis $(\psi_{A_1},\psi_{B_1},\psi_{B_2},\psi_{A_2})^T$, $v_f \approx 10^6$ m/s is the Fermi velocity, $\pi = k_x + i k_y$, and $\boldsymbol{k} = (k_x,k_y)$ denotes the crystal momentum. The electrostatic potential $V_0$ shifts all bands uniformly, while $V_1$ introduces a layer-asymmetric potential that opens a gap between the middle bands~\cite{Oostinga2007Gate}. For numerical convenience and stability, we perform all calculations in dimensionless units by introducing a characteristic length scale $l_0 = 100\,\mathrm{nm}$ and the associated energy scale $E_0 = \hbar v_f / l_0 \approx 6.58\,\mathrm{meV}$. This rescaling reduces the number of free parameters and simplifies the numerical implementation. The dimensionless variables are defined through the transformations
\begin{equation}
\begin{aligned}
V_{0,1} &\rightarrow V_{0,1}/E_0, \qquad
E \rightarrow E/E_0, \\
\gamma_1 &\rightarrow \gamma_1/E_0, \qquad
\boldsymbol{k} \rightarrow l_0\,\boldsymbol{k}.
\end{aligned}
\end{equation}
Physical units are restored in the figures by applying the inverse of these transformations, allowing for direct comparison with realistic relevant energy and length scales.

\subsection{Transport channels in the leads}

We consider an electron with energy $E$ incident from the left $N$ region along the $x$ axis. In the configuration shown in Fig.~\ref{fig:1}(c), the system is translationally invariant along the $y$ direction, so that the transverse momentum $k_y$ is conserved. The wavefunction can therefore be written in the partially Fourier-transformed form $\Psi(x,y)=\Phi(x)e^{ik_y y}$, where $\Phi(x)$ is a four-component spinor (See Appendix~\ref{App: A}). Solving the four-band eigenvalue problem in the unmodulated $N$ regions yields the dispersion relation
\begin{equation}
E = s \frac{\gamma_1}{2} + c \sqrt{k^2 + \frac{\gamma_1^2}{4}},
\label{Eq:Epristine}
\end{equation}
where $c=\pm1$ labels the band index and $s=\pm1$ distinguishes the two low-energy branches. For a given energy, Eq.~\eqref{Eq:Epristine} admits two distinct longitudinal wave vectors, denoted $k^{+}$ and $k^{-}$, corresponding to two independent propagating modes in the $N$ regions. These two modes define four transport channels: two non-scattering channels, $T_{+}^{+}: k^{+}\!\rightarrow k^{+}$ and $T_{-}^{-}: k^{-}\!\rightarrow k^{-}$, and two inter-mode scattering channels, $T_{-}^{+}: k^{+}\!\rightarrow k^{-}$ and $T_{+}^{-}: k^{-}\!\rightarrow k^{+}$, as illustrated in Fig.~\ref{fig:1}(d). Real values of $k^{\pm}$ correspond to propagating states, while complex values describe evanescent modes. In the unmodulated $N$ regions only propagating solutions are allowed, whereas in the modulated $S$ region propagating and evanescent modes generally coexist and jointly determine the scattering process~\cite{Ando1991Quantum,Wu1991Quantum}.

The transmission and reflection amplitudes associated with each channel are obtained using a mode-resolved transfer-matrix approach within a full four-band model~\cite{barbier2010kronig}. Because the two propagating modes carry different group velocities, transmission probabilities are evaluated from the corresponding current densities. The complete derivation of the eigenmodes, matching conditions at the interfaces, construction of the scattering matrix, and explicit expressions for the transmission and reflection coefficients are provided in Appendix~\ref{App: A}.

\subsection{Uniform Strain in BG}

In real space, uniform uniaxial strain modifies the nearest-neighbor bond vectors of the BG lattice according to
$\boldsymbol{\delta}_\zeta = (I + \mathcal{E})\boldsymbol{\delta}^{(0)}_\zeta$
($\zeta=1,2,3$), where the unstrained vectors are $\boldsymbol{\delta}^{(0)}_1 = a(\sqrt{3},1)/2$,
$\boldsymbol{\delta}^{(0)}_2 = a(-\sqrt{3},1)/2$, and $\boldsymbol{\delta}^{(0)}_3 = a(0,-1)$. Here $a=1.42$~\AA\ is the equilibrium carbon-carbon bond length, and $\mathcal{E}$ is the uniaxial strain tensor~\cite{farokhnezhad2017strain}
\begin{equation}
\mathcal{E} = \epsilon
\begin{pmatrix}
\cos^2\theta - \nu \sin^2\theta &
(1+\nu)\cos\theta\sin\theta \\
(1+\nu)\cos\theta\sin\theta &
\sin^2\theta - \nu \cos^2\theta
\end{pmatrix},
\end{equation}
where $\epsilon$ is the strain magnitude, $\theta$ is the strain direction with respect to the lattice axes, and $\nu=0.14$ is the Poisson ratio of graphene~\cite{pellegrino2012resonant}. Strain affects the electronic structure of BG through two related mechanisms. First, lattice deformation alters the geometry of reciprocal space, leading to a rescaling and distortion of isoenergetic contours even in the absence of hopping renormalization. Second, strain modifies the intralayer hopping amplitudes through bond-length variations, resulting in a renormalization of the Fermi velocity~\cite{Oliva2013Understanding,Oliva2015Generalizing,Naumis2017Electronic}. In realistic situations both effects are present simultaneously.

To first order in the strain magnitude, each graphene layer retains a Dirac-like structure, but the deformation shifts the valleys away from their unstrained positions at $\pm K$ and introduces an anisotropic Fermi velocity~\cite{castro2009electronic,pereira2009tight,pellegrino2012resonant}. Following Ref.~\cite{pellegrino2011transport}, these effects can be incorporated into
Eq.~\eqref{eq: MainHamiltonian} by replacing $\pi^\pm$ with $\Pi^\pm$, where
\begin{equation}
\Pi^\pm = (1-\lambda_x \epsilon)\, q_x \pm i (1-\lambda_y \epsilon)\, q_y
\end{equation}
accounts for the strain-induced anisotropy of the Fermi velocity. The strain coefficients are
$\lambda_x = 2\kappa$ and $\lambda_y = -2\kappa\nu$, with $\kappa = \kappa_0 - 1/2$ and
$\kappa_0 \approx 1.6$ a constant related to the logarithmic
derivative of the nearest-neighbor hopping~\footnote{for details of the derivation for the monolayer case please refer to Ref.~\cite{pellegrino2011transport}}. The momentum $\boldsymbol{q}=(q_x,q_y)$ is measured relative to the strain-shifted Dirac point,
$\boldsymbol{q} = \boldsymbol{p} - \boldsymbol{q}_D$, whose position is
\begin{equation}
\boldsymbol{q}_D a =
\left(
\kappa_0 \epsilon (1+\nu) \cos 2\theta,\;
-\kappa_0 \epsilon (1+\nu) \sin 2\theta
\right),
\label{eq:Diracshift}
\end{equation}
and $\boldsymbol{p}=\{k_x,k_y\}$. The resulting four bands energy spectrum with strain is then given by
\begin{equation}
E = V_0 \pm \varepsilon_{\pm}(q),
\end{equation}
with electron and hole bands given by
\begin{equation}
\varepsilon_{\pm}(q)=
\sqrt{
q^2 + V_1^2 + \frac{\gamma_1^2}{2}
\pm
\frac{1}{2}\sqrt{\gamma_1^4 + 4q^2\left(\gamma_1^2+4V_1^2\right)}
}.
\end{equation}
where $q^2 = (1-\lambda_x \epsilon)^2 q_x^2 + (1-\lambda_y \epsilon)^2 q_y^2$. Although strain acts on the Dirac cones of the individual layers, the system has a quadratic low-energy band structure~\cite{mccann2006landau}, whose band-touching point is shifted in momentum space by the applied deformation, as illustrated in Fig.~\ref{fig:Bandas}.

In the following, we consider the combined effects of an electrostatic potential, mass gap and a uniform strain modulation. The transverse momentum $k_y$ remains a good quantum number in both $N$ and $S$ regions. The explicit form of the wave functions and the derivation of the transmission and reflection coefficients are given in Appendix~\ref{App: A}. The zero-temperature conductance is obtained from the mode-resolved transmission probabilities~\cite{Liu2022Pure}
\begin{equation}
G(E) = G_0
\sum_{s,s'}
\int_{-\pi/2}^{\pi/2}
T^s_{s'}(E,\phi)\,
\cos\phi\, d\phi,
\label{eq:conductance}
\end{equation}
where $T^s_{s'}$ is the transmission coefficient from mode $s$ to mode $s'$ with $G_0 = 2e^2/h$ the conductance quantum and $\phi$ is the incident angle.

\begin{figure*}
  \centering 
  \includegraphics[scale=0.65]{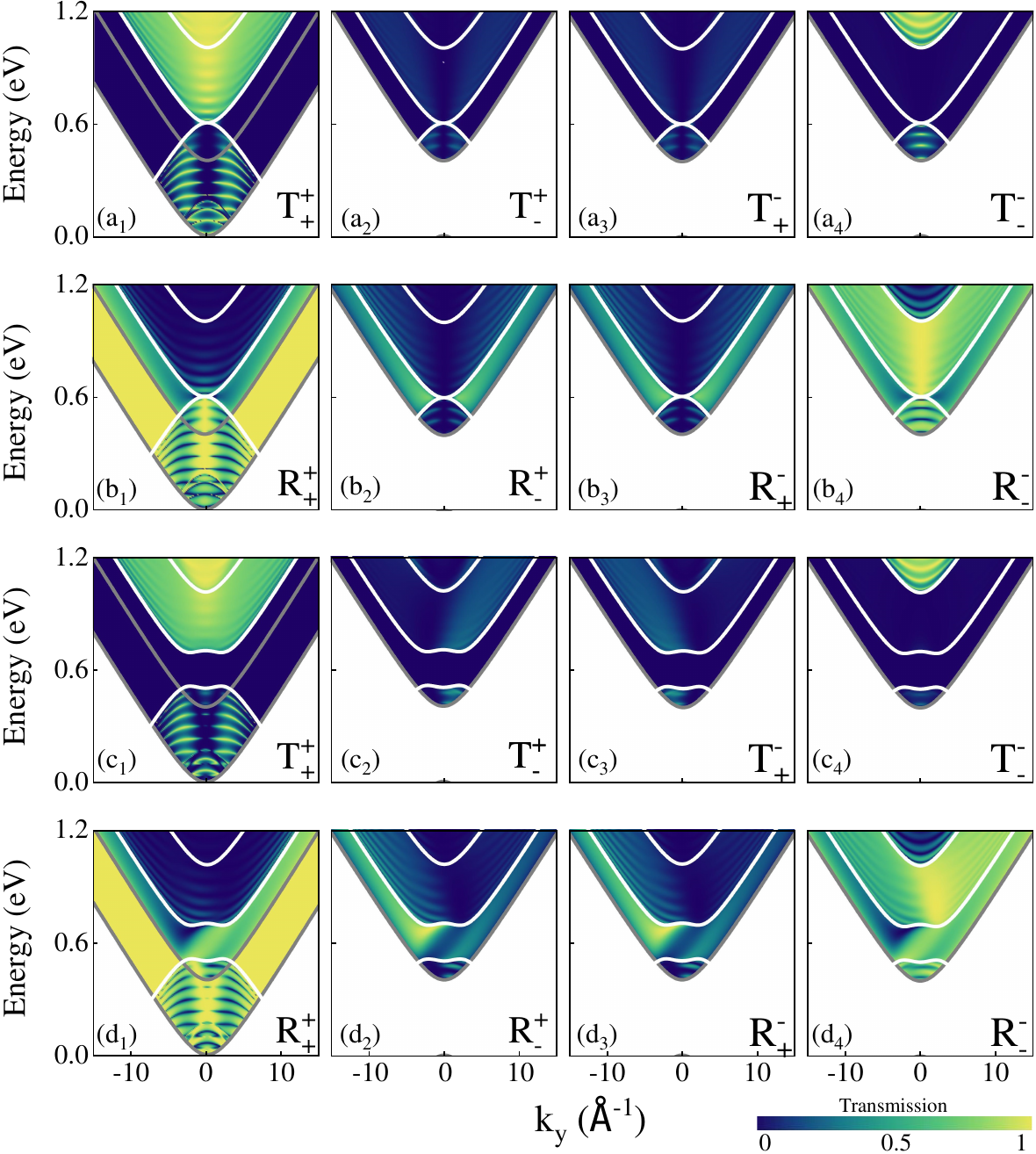}
  \caption{Variation of four-band tunneling transmission probabilities with energy $E$ and k$_{y}$ for a modulation region of length $L=25$ nm and zero strain ($\epsilon=0\%$). Each panel illustrate the corresponding transmission and reflection amplitudes. First two rows are for $V_{0}$=0.6 eV and $V_1$=0., two bottom rows are for $V_{0}$=0.6 eV and $V_1$=0.1 eV. Gray (white) lines represents the boundaries in the $E-k_{y}$ plane for the N (S) region, respectively.}
  \label{fig:Tbarrier}
\end{figure*}

\section{RESULTS}

\subsection{Device with an electrostatic potential}

We first consider the reference case of a device in which the strained region in Fig.~\ref{fig:1} is replaced by a purely electrostatic rectangular barrier~\cite{,Katsnelson2006Chiral}. While this system has been widely investigated~\cite{van2013four,barbier2009bilayer,Van2016Transport}, the transport behavior is often discussed without a unified, mode-resolved classification of the different propagation regimes. In this subsection, we make this structure explicit, providing a detailed interpretation of the transport maps that will serve as a reference for the effects of interlayer bias and strain discussed below.

For a simple barrier, the four bands in the central region are rigidly shifted by the potential $V_0$, as illustrated in Fig.~\ref{fig:1}(b). Fig.~\ref{fig:Tbarrier} shows the resulting transmission and reflection probabilities as functions of the incident energy $E$ and transverse momentum $k_y$. The first two rows correspond to $V_0 = 0.6$~eV without interlayer bias, while the bottom two rows show the case $V_0 = 0.6$~eV and $V_1 = 0.1$~eV. For other choices of electrostatic and bias potentials, the band edges and mode boundaries are shifted in energy and momentum space, but the qualitative transport behavior and the underlying physical mechanisms remain the same within the corresponding parameter ranges. In Fig.~\ref{fig:Tbarrier}, gray and white lines mark the boundaries separating the transport modes in the $N$ and $S$ regions, respectively. These boundaries define distinct transport regimes depending on whether the modes $k^\pm$ and $q^\pm$ are propagating or evanescent. For $V_1=0$, inversion symmetry is preserved, implying
$\mathrm{T}_{+}^{-}=\mathrm{T}_{-}^{+}$ and
$\mathrm{R}_{+}^{-}=\mathrm{R}_{-}^{+}$~\cite{van2013four}.

The schematic shown in Fig.~\ref{fig:1}(b) defines a region-by-region classification of the $(E,k_y)$ plane based on whether the modes available in the leads and in the barrier are propagating or evanescent. The blue and red boundaries correspond to the mode thresholds in the $N$ and $S$ regions, respectively, and they are the same boundaries overlaid on the transmission maps in Fig.~\ref{fig:Tbarrier}. Throughout this section, we use this classification as a reference to organize and interpret the transport behavior shown in Fig.~\ref{fig:Tbarrier}.

In regions~1 and~2, the modes $k^{+}$, $q^{+}$, and $q^{-}$ are propagating, while $k^{-}$ is evanescent. In region~1, illustrated in Fig.~\ref{fig:Tbarrier}(a$_1$), the $\mathrm{T}_{+}^{+}$ channel exhibits pronounced transmission peaks associated with perfect resonances inside the barrier~\cite{Liu2026Mode}. In region~2, despite the presence of propagating states within the barrier, transmission is strongly suppressed at normal incidence ($k_y=0$), reflecting symmetry-imposed decoupling, or cloaking, between the incident modes and the internal barrier states~\cite{Gu2011Chirality}. As a consequence, these internal states become effectively invisible to transport~\cite{Gu2011Chirality,Lee2016Evidence,Katsnelson2006Chiral}, while the remaining transmission channels do not contribute because the $k^{-}$ mode is evanescent in this energy range.

Region~3 is of particular interest. In this regime, both $k^{\pm}$ modes are propagating in the leads, while the $q^{-}$ mode inside the barrier becomes evanescent. Transmission through this region therefore requires coupling either to the evanescent mode $q^{-}$ or to the propagating mode $q^{+}$. At normal incidence, the mode $q^{+}$ is cloaked with respect to $k^{+}$, and the mode $q^{-}$ is cloaked with respect to $k^{-}$~\cite{Liu2026Mode}. As a result, transmission is strongly suppressed at normal incidence and only weak transmission appears away from it. At the same time, the $\mathrm{T}_{-}^{-}$ channel exhibits pronounced resonant features with relatively large transmission. This enhancement arises because the $q^{+}$ mode becomes accessible to the $k^{-}$ channel.

In region~4, only the $k^{+}$ mode remains propagating in the leads. Transmission therefore occurs through tunneling via the evanescent mode $q^{-}$ and through non-normal incidence coupling to the $q^{+}$ mode, which is cloaked at normal incidence. As a consequence, $\mathrm{T}_{+}^{+}$ is strongly reduced and reflection is enhanced, while the remaining transmission channels are suppressed. In region~5, transport proceeds solely via evanescent tunneling, resulting in weak transmission across all channels.

Regions~6 and~7 highlight distinctive features of the four-band description, see also Fig.~\ref{fig:1}(b). In these regimes, even though the incident energy lies above the electrostatic barrier height $V_0$, transport through the $k^{-}$ mode remains inhibited because the high-energy branch acts as an effective barrier of height $V_0+\gamma_1$ for this channel.

In region~6, transmission in the $\mathrm{T}_{+}^{+}$ channel shows clear Fabry-P\'erot resonances at normal incidence, reflecting propagation above a barrier of height $V_0$ through the $k^{+}$ mode. Away from normal incidence, transmission in this channel remains large, while channels involving the $k^{-}$ mode are strongly suppressed. In particular, $\mathrm{T}_{-}^{-}$ is nearly zero because transport proceeds via an evanescent mode, a direct consequence of the effective barrier $V_0+\gamma_1$ experienced by the $k^{-}$ sector. This suppression is further reinforced by the fact that, in this region, the $q^{+}$ mode is cloaked at normal incidence for the $k^{-}$ channel. Small transmission appears only away from normal incidence due to mode mixing.

\begin{figure*}
  \centering
  \includegraphics[scale=0.65]{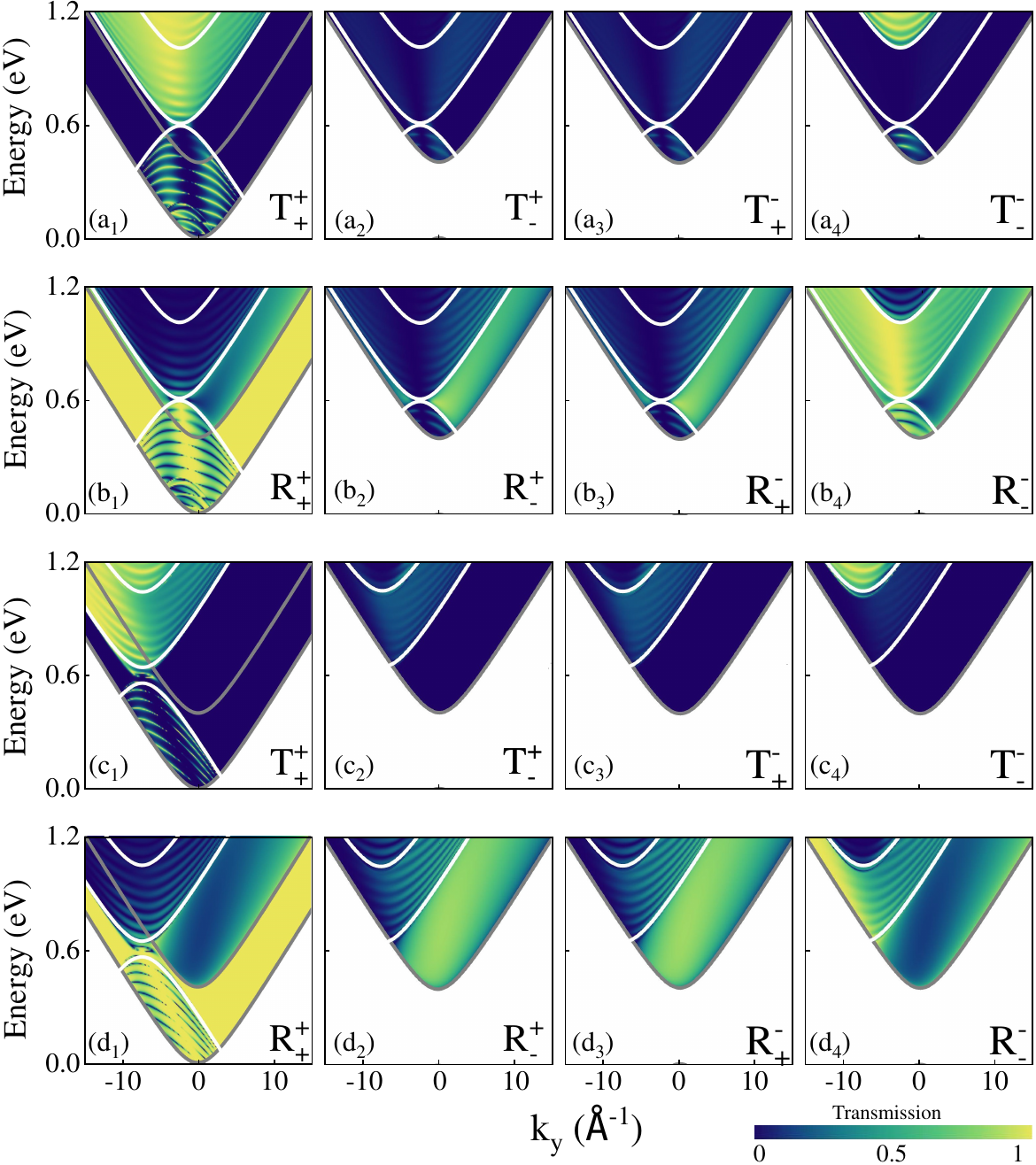}
  \caption{Transmission and reflection coefficients in strained BG as functions of energy $E$ and $k_{y}$ and uniform barrier height $V_{0}$=0.6 eV, an interlayer bias $V_1$=0, and strain direction $\theta$=0.2$\pi$. The strain magnitude is set to (a$_{1}$)-(a$_{4}$) and (b$_{1}$)-(b$_{4}$): $\epsilon$=2 $\%$; (c$_{1}$)-(c$_{4}$) and (d$_{1}$)-(d$_{4}$): $\epsilon$=6$\%$.}
  \label{fig:TStrain}
\end{figure*}

Region~7 presents a different situation. Here, transmission in both the $\mathrm{T}_{+}^{+}$ and $\mathrm{T}_{-}^{-}$ channels displays a Schr\"odinger-like behavior. In the $\mathrm{T}_{+}^{+}$ channel, this occurs because the energy lies above the electrostatic barrier of height $V_0$. In the $\mathrm{T}_{-}^{-}$ channel, the same energy lies above the effective barrier of height $V_0+\gamma_1$. As a result, both channels support propagating modes, leading to a large transmission. The enhancement of transmission in the high-energy channel therefore provides a direct transport signature of the interlayer coupling strength~$\gamma_1$. These regimes cannot be captured within effective two-band descriptions, as the appearance of the effective barrier $V_0+\gamma_1$ and the associated transmission features arise from the explicit inclusion of the high-energy bands in the four-band model.

Taken together, the purely electrostatic case provides a clear reference for mode-resolved transport in bilayer graphene junctions. Transmission is governed by the availability of propagating modes in the barrier region and by symmetry-imposed selection rules that control their coupling to incident states~\cite{van2013four,barbier2009bilayer}. Fabry-P\'erot resonances originate from phase coherence within non-decoupled channels, while the suppression of transmission at normal incidence reflects the persistence of symmetry-protected mode decoupling~\cite{Gu2011Chirality,Huang2025Evanescent,Liu2026Mode}. These features define the baseline transport behavior against which modifications of the junction can be quantitatively assessed.

\begin{figure}
  \centering
  \includegraphics[width =1 \columnwidth]{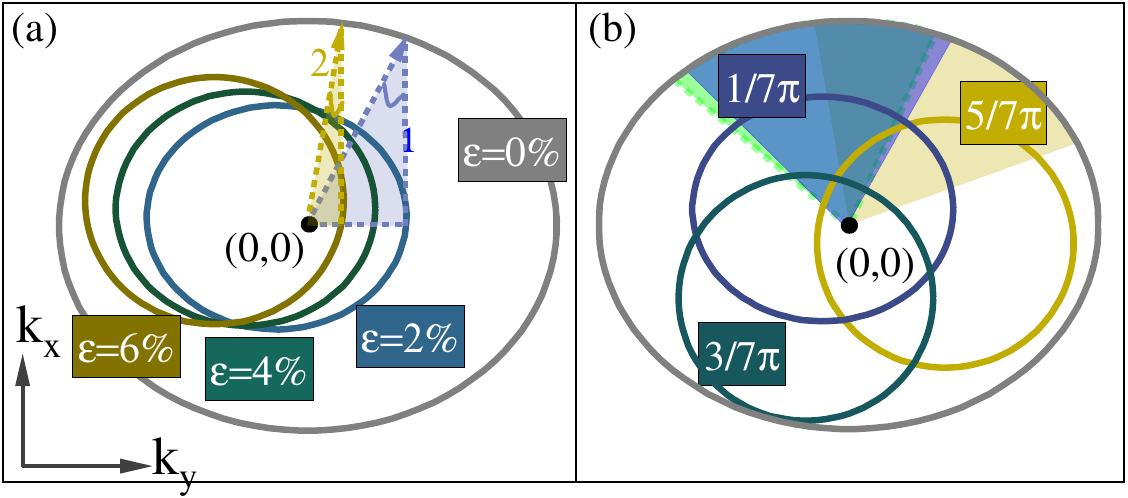}
  \caption{(a) and (b) isoenergetic contours of the strained regions (colored circles) and the unstrained regions (gray circles) under varying magnitudes and directions of strain, respectively. In (a), $\phi_{1}$ and $\phi_{2}$ represent the maximum incident angles permitting transmission at a strain angle of $0.2\pi$ and an incident energy of $E=1.1$ eV, corresponding to strain magnitudes of 2$\%$ and 6$\%$, respectively. In (b), the shaded area illustrates how the range of incident angles allowing transmission varies with an increase in the strain angle at a fixed strain magnitude of $2\%$.}
  \label{fig:Contours}
\end{figure}

\subsection{Effects of a Perpendicular Electric Field}
We now consider a bilayer graphene device in the presence of an electrostatic barrier of height $V_0=0.6$~eV together with a perpendicular electric field $V_1=0.1$~eV, which opens a gap between the two middle bands, as illustrated in Fig.~\ref{fig:Bandas}(d). The corresponding transmission and reflection probabilities are shown in the third and fourth rows of Fig.~\ref{fig:Tbarrier}, respectively. As in the purely electrostatic case, the barrier rigidly shifts the bands in the central region. In addition, the perpendicular electric field breaks inversion symmetry in the $S$ region, induces a gap, and mixes modes that are symmetry-decoupled in the gapless case (see Appendix~\ref{App: A}).

At low energies, corresponding to regions~1 and~2 of Fig.~\ref{fig:1}(b), the transmission retains several qualitative features observed for $V_1=0$. However, additional transmission peaks appear near normal incidence. These features indicate a partial breakdown of the cloaking mechanism, originating from the mass-induced mixing between modes that are decoupled in the absence of an interlayer bias~\cite{Nilsson2007Transmission,van2013four}. As a result, modes inside the barrier become weakly coupled to incident propagating modes, allowing finite transmission even close to $k_y=0$. The perpendicular field also slightly modifies the oscillation period, reflecting the change in the barrier height caused by the gap opening.

In region~3, transport follows a similar behavior as in the pure barrier case. While the non-scattering channels $\mathrm{T}_{+}^{+}$ and $\mathrm{T}_{-}^{-}$ remain symmetric because propagation occurs within the same mode, the breaking of inversion symmetry leads to an asymmetry between the scattering channels, such that $\mathrm{T}_{-}^{+} \neq \mathrm{T}_{+}^{-}$, as shown in Fig.~\ref{fig:Tbarrier}(c$_2$) and Fig.~\ref{fig:Tbarrier}(c$_3$). This asymmetry originates from the layer-dependent potential introduced by the perpendicular electric field~\cite{Nilsson2007Transmission,van2013four}.

The behavior in the remaining regions follows the same mode-based classification as in the absence of the electric field. The essential effect of the perpendicular field is therefore twofold: it lifts the symmetry protection responsible for cloaking by mixing internal modes, and it breaks the $k_y\rightarrow -k_y$ symmetry of mode-changing processes. These effects clearly modify the detailed structure of the transmission maps.

\begin{figure*}
  \centering
  \includegraphics[scale=0.35]{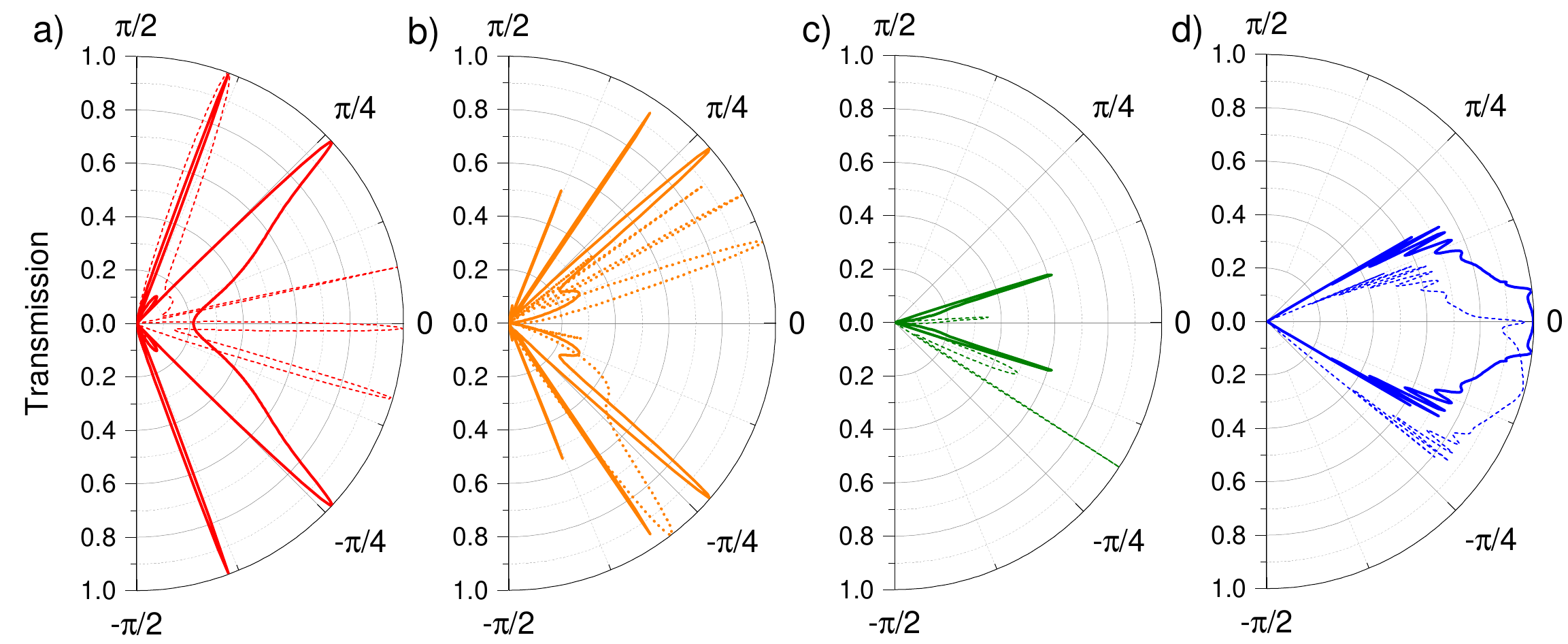}
  \caption{Transmission probability in the channel $T^+_+$ through a single uniform electrostatic barrier with height $V_0 = 0.6$~eV and zero interlayer bias ($V_1 = 0$) as a function of the incident angle. Panels a)-d) correspond to incident energies $E = 0.1$, $0.3$, $0.5$, and $1.1$~eV, respectively. Solid lines show the unstrained case ($\epsilon = 0$), while dashed lines correspond to homogeneous strain with magnitude $\epsilon = 2\%$ and direction $\theta = 2\pi$.}
  \label{fig:AngleDependence}
\end{figure*}

\subsection{Effects of Strain}

We now consider the effect of homogeneous in-plane strain on the transport properties. As discussed above, strain shifts the quadratic band-touching point in momentum space and renormalizes the Fermi velocity~\cite{deJuan2012space,Oliva2013Understanding,Oliva2015Generalizing}. The resulting strain-induced modifications of the BG band structure are illustrated in Fig.~\ref{fig:Bandas}(b) and Fig.~\ref{fig:Bandas}(c). Figure~\ref{fig:TStrain} shows the transmission and reflection probabilities as functions of the transverse momentum $k_y$ and the incident energy $E$ for all four transmission channels under different strain configurations. The first two rows correspond to a strain magnitude $\epsilon=2\%$, while the third and fourth rows show the results for $\epsilon=6\%$. In all cases, the strain direction is fixed to $\theta=0.2\pi$. Gray (white) lines denote the boundaries separating propagating and evanescent modes in the unstrained $N$ (strained $S$) regions.

As illustrated in Fig.~\ref{fig:Bandas}(b), strain displaces the quadratic band-touching points, leading to a corresponding shift of the mode boundaries in the $S$ region. This displacement produces an apparent gap in the $(E,k_y)$ representation, which is not physical but instead reflects the fact that the spectra are evaluated along the fixed cut $k_x=0$, rather than along the direction passing through the strain-shifted band-touching point.

In contrast to the case with a perpendicular electric field, strain induces asymmetry with respect to $k_y$ in both scattering and non-scattering channels. This reflects the geometric nature of strain, which breaks the $k_y \rightarrow -k_y$ symmetry~\footnote{Although uniform strain breaks the symmetry $k_y \rightarrow -k_y$ of the dispersion relation, the Hamiltonian remains translationally invariant along the $y$ direction; therefore, $k_y$ is still conserved and remains a good quantum number.
}. While resonant features persist under strain, they become asymmetric in momentum space. This behavior is illustrated in Fig.~\ref{fig:AngleDependence}, which shows the angular dependence of the transmission in the $\mathrm{T}_{+}^{+}$ channel for several incident energies. Compared to the unstrained case, strain suppresses the overall transmission and shifts the angular positions of the transmission maxima, while preserving the underlying interference structure.

A comparison between the results for $\epsilon=2\%$ and $\epsilon=6\%$ shows that increasing the strain magnitude leads to a pronounced suppression of the transmission across all channels. This suppression originates from the progressive shift of the strained bands, which reduces the overlap between propagating states in the $N$ and $S$ regions and thereby decreases the number of available transport channels. As a result, strain acts as an efficient control parameter for tuning both the magnitude and angular selectivity of ballistic transport in BG junctions.

\subsection{Iso-energetic interpretation of the transmission}

The sensitivity of the transmission to strain can be understood as a direct consequence of the strain-induced displacement of the quadratic band-touching points, which reduces the overlap between states available for transport across the junction. As illustrated in Fig.~\ref{fig:Contours}, an incident carrier with energy $E$ in the unstrained $N$ region is characterized by an isoenergetic contour (gray curve) defining the set of allowed wave vectors. For transmission into the strained $S$ region, energy conservation requires the transmitted state to lie on the corresponding isoenergetic contour at the same energy. Because strain deforms and shifts the band structure, the isoenergetic contours in the $S$ region differ in both shape and position, leading to a momentum-space mismatch between the two regions.

The isoenergetic contours for the $N$ and $S$ regions under different strain conditions are shown in Fig.~\ref{fig:Contours}. In these plots, the coordinate axes are rotated such that the $k_y$ direction lies along the horizontal axis. As the strain magnitude increases, the contours in the strained region are displaced according to the shift of the quadratic band-touching points given by Eq.~\ref{eq:Diracshift}. This displacement breaks the symmetry of the overlap between the $N$ and $S$ contours with respect to normal incidence ($k_y=0$), providing a geometric origin for the transmission asymmetry observed in Fig.~\ref{fig:TStrain}.

In addition to inducing asymmetry, strain also suppresses the overall transmission by narrowing the range of incident angles that support propagating solutions in the $S$ region. This effect can be visualized by considering the incident angle $\varphi=\arctan(k_y/k_x)$ at fixed energy. For a given strain configuration, transmission is allowed only for values of $k_y$ for which a vertical line of constant $k_y$ intersects the strained isoenergetic contour. As illustrated for $\epsilon=2\%$ in Fig.~\ref{fig:Contours}, this condition defines a finite angular window $(-\varphi_1,\varphi_1)$. As the strain magnitude increases from $2\%$ to $4\%$ and $6\%$, the maximum transverse momentum allowed for propagation in the $S$ region decreases from approximately $0.08~\text{\AA}^{-1}$ to about $0.03~\text{\AA}^{-1}$, corresponding to a reduction of the maximum incident angle from $\varphi_1\approx 86^\circ$ to $\varphi_2\approx 56^\circ$. This progressive reduction explains the strong suppression of transmission observed at larger strain.

A related but distinct effect arises when the strain direction $\theta$ is varied, as shown in Fig.~\ref{fig:Contours}(b). In this case, the isoenergetic contours retain their size and shape but rotate and shift in momentum space. This behavior follows from the elliptical form of the contours,
\begin{equation}
\frac{(k_x-q_{Dx})^2}{v_y^2}+\frac{(k_y-q_{Dy})^2}{v_x^2}-\frac{\mathcal{C}_0}{v_x^2 v_y^2}
=0,
\end{equation}
where $\mathcal{C}_0=V_1^2+E^2+\sqrt{4V_1^2E^2+\gamma_1^2(E^2-V_1^2)}$, and the strain-induced shift $\boldsymbol{q}_D$ is given by Eq.~\ref{eq:Diracshift}. Changing $\theta$ therefore shifts the center of the ellipse without modifying the lengths of its principal axes.

As a result, the width of the angular transmission window remains approximately constant, while its center is displaced in momentum space. The primary role of the strain direction is thus to steer the angular interval over which tunneling is most efficient, providing directional control of electron transport without significantly altering the intrinsic band structure or the overall angular range that supports transmission.

\subsection{Strain-Dependent Cloaking Effect}

Having established how homogeneous strain reshapes the isoenergetic contours and redistributes the angular transmission, we now examine how these geometric modifications affect the cloaking mechanism that governs transport at normal incidence~\cite{Gu2011Chirality,Liu2026Mode}. 

At normal incidence, the BG Hamiltonian supports propagating states in the $N$ regions and both propagating and evanescent solutions inside the barrier in the $S$ region. Due to symmetry constraints, a subset of these internal modes remains exactly decoupled from the incident propagating channels and therefore does not contribute to transmission, even though these solutions exist at energies accessible to transport~\cite{Liu2026Mode}. As a result, the corresponding internal modes become effectively invisible to incident carriers, giving rise to the cloaking effect, often discussed in the literature under the label of anti-Klein tunneling~\cite{Katsnelson2006Chiral,Varlet2014Fabry,AgrawalGarg2012Reversal,Chen2009Design}. From a microscopic perspective, this decoupling arises because, for $V_1=0$ and vanishing transverse momentum inside the barrier ($q_y=0$), the four coupled differential equations governing transport in region $S$ become exactly separable into two independent sectors. This symmetry-protected separation implies that only one of the two internal branches couples to the incoming lead modes, while the other remains strictly orthogonal and thus cloaked~\cite{Huang2025Evanescent}. In the absence of strain, the condition $q_y=0$ coincides with normal incidence in the leads ($k_y=0$). In contrast, under homogeneous strain, the same decoupling condition is shifted to finite transverse momentum according to $q_y = k_y - q_{Dy} = 0$, corresponding to oblique incidence in the leads with $k_y = q_{Dy}$, where $q_{Dy}$ denotes the $y$ component of the Dirac-point shift defined in Eq.~\eqref{eq:Diracshift}.

\begin{figure*}
  \centering
  \includegraphics[scale=0.40]{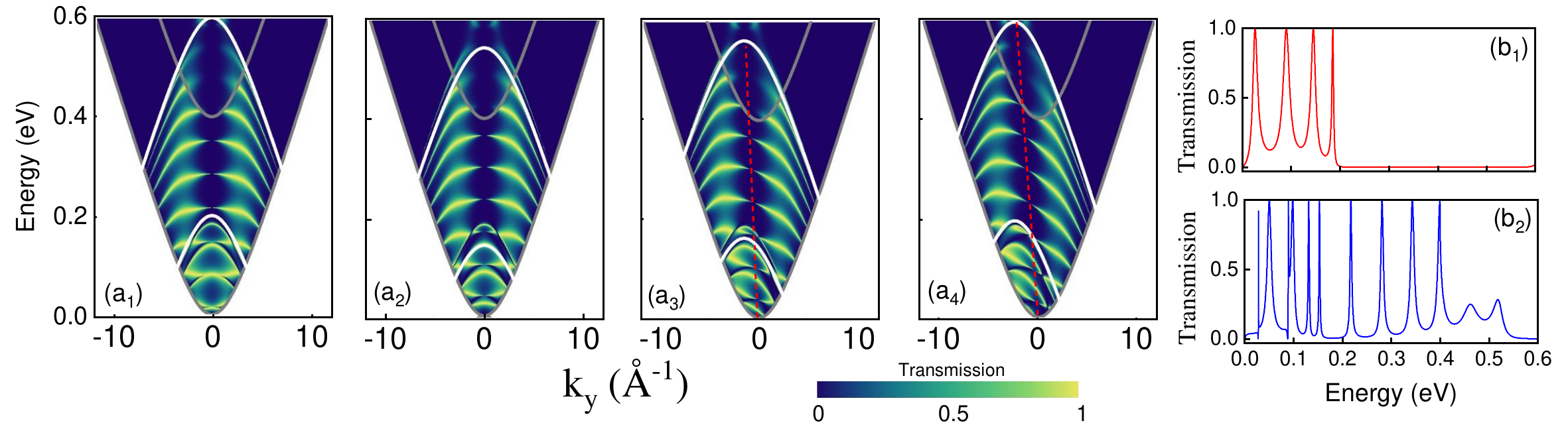}
  \caption{Transmission in the channel $\text{T}_+^+$ with: (a$_1$) $\epsilon = 0$, (a$_2$) $\epsilon = 2 \%$ and $\phi = 0 \pi$, (a$_3$) $\epsilon = 2 \%$ and $\phi = 0.1 \pi$ and (a$_4$) $\epsilon = 2 \%$ and $\phi = 0.2 \pi$. The transmission for normal incidence $k_y=0$ is shown in (b$_1$) and (b$_2$) for the panels (a$_2$) and (a$_3$), respectively. In all figures we set $V_{0}$=0.6 eV and $V_1 =0$. Red dashed lines in (a$_3$) and (a$_4$) roughly illustrate the shift of the zero transmission cloaked points.}
  \label{fig:cloack}
\end{figure*}

Fig.~\ref{fig:cloack} illustrates this behavior through the transmission in the $\mathrm{T}_{+}^{+}$ channel for different strain configurations. In the unstrained case, $\epsilon=0$, shown in Fig.~\ref{fig:cloack}(a$_1$), region~1 (see Fig.~\ref{fig:1}(b)) exhibits a sequence of Fabry-Pérot oscillations associated with the propagation process $k^{+}\rightarrow q^{-}\rightarrow k^{+}$. As the incident energy increases, the transmission at normal incidence ($k_y=0$) becomes strongly suppressed, signaling the onset of the cloaking regime. For oblique incidence ($k_y\neq 0$) or in the presence of a mass term, transmission is recovered due to symmetry-allowed mode mixing between propagating channels~\cite{Gu2011Chirality}. Remarkably, the application of strain does not eliminate the cloaking mechanism. As shown in Fig.~\ref{fig:cloack}(a$_2$) for $\epsilon=2\%$ and $\theta=0$, the suppression of transmission persists at normal incidence. However, varying the strain direction shifts the cloaking condition away from $k_y=0$ toward a finite transverse momentum determined by the strain-induced momentum shift given by $q_{Dy}$. This displacement is clearly visible in Fig.~\ref{fig:cloack}(a$_3$) and Fig.~\ref{fig:cloack}(a$_4$) for $\theta=0.1\pi$ and $\theta=0.2\pi$, respectively.

These results indicates that homogeneous strain does not destroy the symmetry-imposed mode decoupling responsible for cloaking, but instead shifts its condition in momentum space. This effect can also be obtained by setting $q_y=0$ (and $V_1=0$) which decouples the differential equations in Appendix~\ref{App: A}. Cloaking therefore remains a robust feature of BG transport, while becoming continuously tunable through strain-induced geometric deformation of the band structure. This provides a controlled route to shifting the visibility of otherwise decoupled internal modes without introducing additional scattering mechanisms.

\begin{figure}
  \centering
  \includegraphics[width =1 \columnwidth]{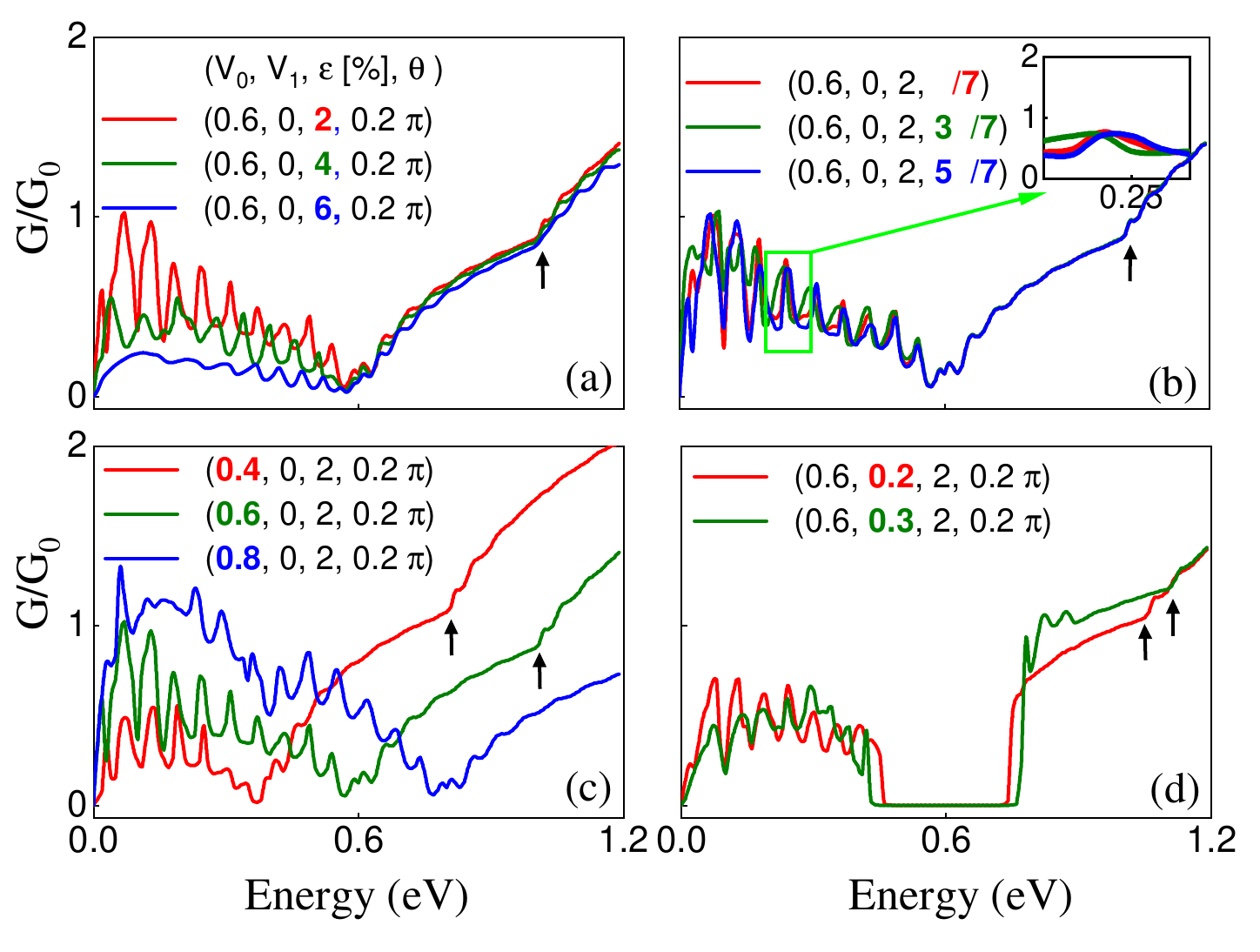}
  \caption{Normalized conductance $G/G_{0}$ as a function of the incident energy for different structural and strain configurations. (a) Variation with strain magnitude: $V_{0}=0.6$ eV, $V_1=0$, $\theta=0.2\pi$, and $\epsilon=2\%, 4\%, 6\%$; (b) Variation with strain direction: $V_{0}=0.6$ eV, $V_1=0$, $\epsilon=2\%$, and $\theta=\pi/7, 3\pi/7, 5\pi/7$; (c) Variation with barrier height: $\epsilon=2\%$, $\theta=0.2\pi$, and $V_0=0.4, 0.6, 0.8$ eV; (d) Variation with bias potential: $V_{0}=0.6$ eV, $\epsilon=2\%$, $\theta=0.2\pi$, and $V_1=0.2, 0.3$ eV. Solid and dashed curves correspond to the different parameter sets indicated in each panel. Black arrows mark the energy at which the slope of the conductance changes. At this energy, the internal high-energy branch inside the barrier no longer acts as an effective barrier for the $k^{-}$ channel, leading to a pronounced increase in its transmission contribution and, consequently, a visible enhancement of the total conductance.}
  \label{fig:Conductance}
\end{figure}

\subsection{Conductance}

The conductance is obtained by integrating the contributions of all propagating channels over the incident momenta, as defined in Eq.~\ref{eq:conductance}. Figure~\ref{fig:Conductance} shows the resulting conductance as a function of the incident energy for different values of the electrostatic potential $V_{0}$, interlayer bias $V_1$, strain magnitude $\epsilon$, and strain direction $\theta$.

Fig.~\ref{fig:Conductance}(a) illustrates the effect of varying the strain magnitude. The conductance exhibits a pronounced dependence on $\epsilon$: as the strain increases, the conductance is progressively suppressed, with a substantial reduction of the resonant peaks for $E<V_{0}$. This behavior reflects the strain-induced mismatch between the sets of propagating states available in the $N$ and $S$ regions, which reduces the transmission probability after angular integration. For $E>V_{0}$, the conductance curves corresponding to different strain magnitudes become increasingly similar, consistent with the reduced sensitivity of higher-energy carriers to strain or barrier-induced mode mismatch.

The influence of the strain direction is shown in Fig.~\ref{fig:Conductance}(b). Varying $\theta$ from $\pi/7$ to $5\pi/7$ produces nearly identical conductance profiles, with only minor local modifications of the oscillatory structure for $E<V_{0}$, as highlighted in the inset. This weak dependence arises because the strain direction primarily redistributes transmission in angle, or equivalently in $k_y$, while leaving the total transmission strength largely unchanged after integration over all incident momenta. This behavior is fully consistent with the isoenergetic-contour interpretation discussed in Fig.~\ref{fig:Contours}(b).

Fig.~\ref{fig:Conductance}(c) shows the conductance for different barrier heights, ranging from $V_{0}=0.4$~eV to $0.8$~eV. The barrier height controls both the energy position and the spacing of the conductance resonances. For $E<V_{0}$, multiple resonance peaks appear as a consequence of phase-coherent propagation through the barrier region. As $V_{0}$ increases, these resonances shift to higher energies and become more widely spaced, reflecting the increased effective barrier strength.

Once the incident energy exceeds $V_{0}$, the conductance increases smoothly. However, a clear change in slope is observed when $E \approx V_{0}+\gamma_{1}$. This feature originates from the four-band structure: although the energy is already above $V_{0}$, the $k^{-}$ branch still experiences an effective barrier of height $V_{0}+\gamma_{1}$. For $E<V_{0}+\gamma_{1}$, transport through this sector remains suppressed because the corresponding mode is evanescent inside the barrier region. When $E$ exceeds $V_{0}+\gamma_{1}$, the $k^{-}$ mode becomes propagating, and an additional transport channel is activated. Its contribution is then incorporated into the angular integration, producing a marked enhancement of the total conductance and a visible increase in the slope of the curves. 

This threshold provides a direct transport signature of the high-energy band. Since the position of this feature is determined by $V_{0}+\gamma_{1}$, systematic measurements as a function of the calibrated barrier height $V_{0}$ allow one, in principle, to extract the interlayer coupling parameter $\gamma_{1}$ from conductance data.

Finally, Fig.~\ref{fig:Conductance}(d) illustrates the effect of the interlayer bias. As $V_1$ increases, the conductance is strongly suppressed and can nearly vanish within the energy window associated with the bias-induced gap. This suppression follows directly from the layer-asymmetric potential, which reduces the number of available propagating modes in the barrier region. As a result, tunneling across the junction is strongly inhibited in this energy range, leading to a pronounced reduction of the conductance near $V_{0}\pm \frac{1}{2}V_1$. 

In addition, the high-energy threshold discussed above is shifted by the bias. Because the interlayer potential splits the bands symmetrically, the effective onset of the enhanced $k^{-}$ contribution occurs at energies close to $V_0 + \frac{1}{2}V_1 + \gamma_1$. This produces a corresponding displacement of the conductance slope change observed in the unbiased case. The clear correlation between the conductance minima, the shifted threshold, and the gap edges indicates that transport measurements across electrostatically defined bilayer graphene junctions provide a direct experimental probe of both the bias-induced gap and the interlayer coupling strength.

\section{CONCLUSIONS}
We have investigated mode-resolved ballistic transport in AB-stacked bilayer graphene across normal-modulated-normal junctions using a four-band low-energy description and a transfer-matrix formalism. By explicitly resolving transmission and reflection between the propagating solutions $k^{\pm}$ in the leads and their counterparts $q^{\pm}$ in the modulated region, we identified distinct transport regimes determined by whether the relevant modes are propagating or evanescent. For purely electrostatic barriers, the transmission maps reproduce Fabry-P\'erot resonances and the suppression of transmission at normal incidence arising from symmetry-imposed mode decoupling. 

Importantly, the four-band structure produces a characteristic threshold at which the high-energy branch inside the barrier becomes propagating. Although the $k^{-}$ channel already exists in the leads, its transmission is strongly suppressed below this threshold because the corresponding internal solution behaves effectively as a barrier. When this internal branch becomes propagating, the transmission weight of the $k^{-}$ channel increases, leading to a visible change in the slope of the conductance at high energies. The position of this feature is determined by the barrier height, the interlayer bias, and the interlayer hopping, providing a direct transport signature of the interlayer coupling. In principle, systematic conductance measurements as a function of the calibrated barrier height $V_0$ and bias $V_1$ allow the extraction of $\gamma_1$ from purely electrical transport data.

A perpendicular interlayer bias breaks inversion symmetry, mixes modes, opens a bias-induced gap, and strongly suppresses the conductance within the corresponding energy window. Homogeneous uniaxial strain modifies transport in a qualitatively different way by shifting and distorting the isoenergetic contours in momentum space. This breaks the $k_y \rightarrow -k_y$ symmetry, redistributes the angular range supporting transmission, and suppresses the conductance for $E<V_0$ as the overlap between states in the leads and in the barrier is reduced. Increasing the strain magnitude narrows the angular transmission window, while varying the strain direction mainly redistributes transmission in momentum space. Importantly, strain preserves the symmetry-protected mode decoupling responsible for cloaking, but shifts its condition away from normal incidence.

Our results are directly relevant to recent Corbino-geometry experiments reporting angularly selective tunneling and suppressed transmission at small incidence angles in bilayer graphene junctions~\cite{Elahi2024Direct}. The observed conductance signatures are interpreted as evidence of symmetry-imposed suppression of head-on transmission and the emergence of finite-angle transmission maxima. The mode-resolved framework developed here provides a microscopic interpretation of these observations, showing how symmetry-driven mode decoupling, internal phase coherence, and band-structure geometry determine the angular transmission profile. In particular, our analysis clarifies how geometric deformations of the band structure, such as those induced by homogeneous strain, shift the conditions for transmission suppression and angular selectivity without removing the underlying decoupling mechanism.

Taken together, electrostatic barriers, interlayer bias, and homogeneous strain provide complementary and independent ways to control energy- and angle-dependent transport in bilayer graphene junctions. Electrostatic gating governs interference and mode availability, interlayer bias opens a gap and modifies mode coupling, and strain reshapes the momentum-space structure while preserving symmetry-based decoupling. The identification of a conductance threshold further demonstrates that multiband effects leave measurable fingerprints in ballistic transport, offering a practical route to probe the interlayer coupling strength. These results provide a unified microscopic framework for interpreting angle-resolved transport experiments and clarify the roles of band structure, symmetry, and geometry in ballistic bilayer graphene devices.

\section*{ACKNOWLEDGMENTS}
We thank Francisco Guinea and Ramon Carrillo-Bastos for useful discussions. D.L acknowledges the support from the China Scholarship Council (CSC) program, project ID: 202406960092, and by the Natural Science Foundation of Shaanxi Province, Grant No.\ 2025JC-YBQN-023. P.A.P acknowledged support from the “Severo Ochoa” Programme for Centres of Excellence in R\&D (CEX2020-001039-S/AEI/10.13039/501100011033) financed by MICIU/AEI/10.13039/501100011033 and from NOVMOMAT, Grant PID2022-142162NB-I00 funded by MCIN/AEI/ 10.13039/501100011033 and, by "ERDF A way of making Europe". P.A.P acknowledges funding by Grant No.\ JSF-24-05-0002 of the Julian Schwinger Foundation for Physics Research\textbf{}
\bibliographystyle{apsrev4-1}

%


\appendix
\section{Transmission and reflection coefficients in the presence of uniform strain, electrostatic and interlayer potentials. }
\label{App: A}
In this section, we obtain a general expression to determine the transport coefficients required to calculate the conductance in the presence of the combined effect of strain, electrostatic potential and interlayer bias~\cite{barbier2009bilayer, pellegrino2011transport}. We consider the propagation of a quasi-particle with energy $E$ and incident angle $\phi$ along the $x$-direction. Because the system is invariant in the $y$-direction, $k_{y}$ is a good quantum number. Therefore, we can write a partial Fourier representation of the wavefunction as $\Psi(x,y)=\Phi (x)e^{i k_{y}y}$, with a four-component wavefunction $\Phi (x)= [\phi_{1}(x), \phi_{2}(x), \phi_{3}(x),\phi_{4}(x) ]^{T}$. By dropping the position dependence in the wavefunctions to simplify the notation and with $k_x \rightarrow -i \hbar\frac{d}{d_x}$, the time-independent Schrödinger equation becomes one-dimensional in the propagation direction, this is, $H \Phi(x)=E \Phi(x)$, which yields a set of coupled differential equations in the presence of strain and external potentials given by

\begin{align}
&\left(-i v_x \frac{d}{dx}-v_x q_{_{Dx}}+iv_y q_{y}\right)\phi_{2}=E_{-}\phi_{1}-\gamma_{1}\phi_{3},\\
&\left(-i v_x \frac{d}{dx}-v_x q_{_{Dx}}-iv_y q_{y}\right)\phi_{1}=E_{-}\phi_{2},\\
&\left(-i v_x \frac{d}{dx}-v_x q_{_{Dx}}-iv_y q_{y}\right)\phi_{4}=E_{+}\phi_{3}-\gamma_{1}\phi_{1},\\
&\left(-i v_x \frac{d}{dx}-v_x q_{_{Dx}}+iv_y q_{y}\right)\phi_{3}=E_{+}\phi_{4}.
\end{align}
where $E_\pm=E-V_0\pm V_{1}$, $v_{x} = 1-\lambda_{x}\epsilon$ and $v_{y} = 1-\lambda_{y}\epsilon$. The above equation can be solved through successive decoupling. In particular, the differential equation for $\phi_{1}$ is given by
\begin{equation}
\left(\frac{d^{2}}{dx^{2}}-2iq_{_{Dx}}\frac{d}{dx}-q_{_{Dx}}^{2}-\frac{v_y^2 q_{y}^{2}}{v_x ^{2}}\right)\phi_{1}=\lambda_{\pm}\phi_{1},
\end{equation}
which is a linear second-order differential equation, with 
\small 
\begin{equation}
\lambda_{\pm}=-\frac{\left( E_-^2 + E_+^2 \pm \sqrt{\left( E_-^2 - E_+^2 \right)^2 + 4 E_- E_+ \gamma_{1}^2} \right)}{2 v_x^2}.
\end{equation}\par
 \normalsize
The plane-wave solutions for $\phi_{1}$ can be written in the form
\begin{equation}
\begin{aligned}
\phi_{1}&=A e^{i(q^{+}+q_{_{Dx}})x}+B e^{-i(q^{+}-q_{_{Dx}})x}\\
&+C e^{i(q^{-}+q_{_{Dx}})x}+D e^{-i(q^{-}-q_{_{Dx}})x}
\end{aligned},
\end{equation}
where the wave vector components $q^{\pm}$ satisfy the relation
\small 
\begin{equation}
\begin{aligned}
&q^{\pm 2} v_{x}^{2}+q_{y}^{2} v_{y}^{2}=\\
&\frac{E_{+}^{2}+E_{-}^{2}\mp \sqrt{(E_{+}^{2}-E_{-}^{2})^{2}+4E_{+}E_{-}\gamma_{1}^{2}}}{2}.
\end{aligned}
\end{equation}
\normalsize \par
We note that for a system without any perturbation, it reduces to 
\begin{equation}
k^{s}=\sqrt{E^{2}+s E \gamma_{1}-k_{y}^{2}},
\end{equation}
where $s=\pm$ is the propagation mode. On the other hand, from Eq. A2, we can write, 
\begin{equation}
\begin{aligned}
\phi_{2}&=d^{+}_{+} A e^{i(q^{+}+q_{_{Dx}})x}+d^{+}_{-} B e^{-i(q^{+}-q_{_{Dx}})x}\\
&+d^{-}_{+}C e^{i(q^{-}+q_{_{Dx}})x}+d^{-}_{-} D e^{-i(q^{-}-q_{_{Dx}})x}
\end{aligned},
\end{equation}
where, $d^{s}_{\pm}=\frac{\pm q^{s}v_x -iq_{y}v_y }{E_{-}}$, and the superscript denotes the propagation mode and the subscript the propagation direction. Substituting $\phi_{1}$ and $\phi_{2}$ into Eq. (A1) yields $\phi_{3}$
\begin{equation}
\begin{aligned}
\phi_{3}&=h^{+} A e^{i(q^{+}+q_{_{Dx}})x}+h^{+}B e^{-i(q^{+}-q_{_{Dx}})x}\\
&+h^{-}C e^{i(q^{-}+q_{_{Dx}})x}+h^{-}D e^{-i(q^{-}-q_{_{Dx}})x}
\end{aligned},
\end{equation}
where $h^{s}=\frac{E_{-}^{2}-v_{x}^{2}(q^{s})^{2}-v_{y}^{2}q_{y}^{2}}{\gamma_{1}E_{-}}$, and 
\begin{equation}
\begin{aligned}
\phi_{4}&=f^{+}_{+}h^{+} A e^{i(q^{+}+q_{_{Dx}})x}+f^{+}_{-}h^{+}B e^{-i(q^{+}-q_{_{Dx}})x}\\
&+f^{-}_{+}h^{-}C e^{i(q^{-}+q_{_{Dx}})x}+f^{-}_{-}h^{-}D e^{-i(q^{-}-q_{_{Dx}})x}
\end{aligned}
\end{equation}
Here, $f^{s}_{\pm}=\frac{\pm v_x q^{s}+iv_y q_{y}}{E_{+}}$. Within the transfer matrix method~\cite{barbier2010kronig}, the wavefunctions in the modulated region can be written as
\begin{equation}
\Phi_{II}(x)=\begin{pmatrix}
 \phi_{1}\\\phi_{2}
 \\ \phi_{3}
 \\\phi_{4}
\end{pmatrix}=\Omega_{II} P_{II}(x) \begin{pmatrix}
 A\\B
 \\C
 \\D
\end{pmatrix},
\end{equation}
with
\begin{equation}
\Omega_{II}=\begin{pmatrix}
  1& 1 &1  &1 \\
  d^{+}_{+}&d^{+}_{-}  &d^{-}_{+}  &d^{-}_{-} \\
 h^{+}& h^{+} &h^{-}  &h^{-} \\
 f^{+}_{+}h^{+} &f^{+}_{-}h^{+}  &f^{-}_{+}h^{-} &f^{-}_{-}h^{-}
\end{pmatrix},
\end{equation}
and
\begin{equation}
P_{II}(x)=\begin{pmatrix}
  e^{iK_{1}x}& 0 &0  &0 \\
 0 & e^{-iK_{2} x} &0  &0 \\
 0 & 0 &e^{iK_{3}x} &0 \\
  0& 0 & 0 &e^{-iK_{4}x}
\end{pmatrix}.
\end{equation}
where in the above equation we have made the replacements, $K_{1}=q^{+}+q_{Dx}$, $K_{2}=q^{+}-q_{Dx}$, $K_{3}=q^{-}+q_{Dx}$, $K_{4}=q^{-}-q_{_{Dx}}$. In the unmodulated region, the eigenstates are directly obtained from Eq. (A13) by setting $\epsilon=0$ and $V_{0}=V_1=0$. In the region N, $\Omega$ can be expressed as
\begin{equation}
\Omega_{\mathrm{I(III)}} =
\begin{pmatrix}
1 & 1 & 1 & 1 \\
l_{+}^{+} & l^{+}_{-} & l^{-}_{+} & l_{-}^{-} \\
j^{+} & j^{+} & j^{-} & j^{-} \\
g_{+}^{+}j^{+} & g^{+}_{-}j^{+} & g^{-}_{+}j^{-} & g_{-}^{-}j^{-}
\end{pmatrix},
\end{equation}
where $l^{s}_{\pm} = \frac{\pm k^{s}_{x} - i k_{y}}{E}$, $j^{s} = \frac{E^{2} - (k^{s}_{x})^{2} - k_{y}^{2}}{E\,\gamma_{1}}$,
and $g^{s}_{\pm} = \frac{\pm k^{s}_{x} + i k_{y}}{E}$. In the 
scattering process, $k_{y}$ is conserved, and its expression is $k^{s}_{y} = \sqrt{E^{2} + s E \gamma_{1}}\,\sin\phi$. In the left region, we have
\begin{equation}
P_{\mathrm{I(III)}}(x) =
\begin{pmatrix}
e^{ i k^{+} x } & 0               & 0               & 0 \\
0               & e^{- i k^{+} x} & 0               & 0 \\
0               & 0               & e^{ i k^{-} x } & 0 \\
0               & 0               & 0               & e^{- i k^{-} x}
\end{pmatrix},
\end{equation}
Thus, the eigenstate in the left region has the following form:
\begin{equation}
\Phi_{\mathrm{I}}(x)=\Omega_{\mathrm{I}} P_{\mathrm{I}}(x)\begin{pmatrix}
 \delta_{s,1}\\r^{s}_{+}
 \\\delta_{s,-1}
 \\r^{s}_{-}
\end{pmatrix}.
\end{equation}\par
 $\delta$ is the Kronecker delta. On the right side, the wave function contains only the transmitted component, with no reflected wave and is given by:
\begin{equation}
\Phi_{\mathrm{III}}(x)=\Omega_{\mathrm{III}} P_{\mathrm{III}}(x)\begin{pmatrix}
 t^{s}_{+}\\0
 \\t^{s}_{-}
 \\0
\end{pmatrix}.
\end{equation}\par
By imposing the continuity of the wave function at each interface, we can relate the coefficients in the source region to those in the  drain region through the total transfer matrix. At $x = 0$, the 
matching between the source region (region I) and the strained region (region II) results in
\begin{equation}
\begin{pmatrix}
 \delta_{s,1} \\[2pt]
 r^{s}_{+} \\[2pt]
 \delta_{s,-1} \\[2pt]
 r^{s}_{-}
\end{pmatrix}
=
P_{\mathrm{I}}^{-1}(0)\,
\Omega_{\mathrm{I}}^{-1}\,
\Omega_{\mathrm{II}}\,
P_{\mathrm{II}}(0)
\begin{pmatrix}
 A \\[2pt]
 B \\[2pt]
 C \\[2pt]
 D
\end{pmatrix}.
\end{equation}\par
At the interface between the strained region and region III, located at $x=L$, we have
\begin{equation}
\begin{pmatrix}
A\\
B\\
C\\
D
\end{pmatrix}
=
P_{\mathrm{II}}^{-1}(L)\,
\Omega_{\mathrm{II}}^{L}\,
\Omega_{\mathrm{III}}\,
P_{\mathrm{III}}(L)
\begin{pmatrix}
t^{s}_{+}\\
0\\
t^{s}_{-}\\
0
\end{pmatrix}.
\end{equation}\par
To simplify the notation, we define the interface transfer matrices between region I and region II, and between region II and region III as:
\begin{equation}
\begin{aligned}
M_{\mathrm{I}\to\mathrm{II}}
&=P_{\mathrm{I}}^{-1}(0)\,
\Omega_{\mathrm{I}}^{-1}\,
\Omega_{\mathrm{II}}\,
P_{\mathrm{II}}(0),\\[4pt]
M_{\mathrm{II}\to\mathrm{III}}
&=P_{\mathrm{II}}^{-1}(L)\,
\Omega_{\mathrm{II}}^{-1}\,
\Omega_{\mathrm{III}}\,
P_{\mathrm{III}}(L),
\end{aligned}
\end{equation}
with these definitions, the relation between the coefficients in the source and drain regions takes the compact form
\begin{equation}
\begin{pmatrix}
\delta_{s,1}\\
r^{s}_{+}\\
\delta_{s,-1}\\
r^{s}_{-}
\end{pmatrix}
=
S(L,R)
\begin{pmatrix}
t^{s}_{+}\\
0\\
t^{s}_{-}\\
0
\end{pmatrix},
\end{equation}
where the total scattering matrix is
\begin{equation}
S(L,R)=M_{\mathrm{I}\to\mathrm{II}}\,M_{\mathrm{II}\to\mathrm{III}}.
\label{eq: Connection}
\end{equation}\par
For convenience, we introduce the common denominator
\begin{equation}
\Delta = S_{11}S_{33}-S_{13}S_{31},
\end{equation}
the transmission amplitudes are then given by
\begin{equation}
\begin{aligned}
t^{+}_{+} &= \frac{S_{33}}{\Delta},\qquad
t^{+}_{-} = -\frac{S_{31}}{\Delta},\\[2pt]
t^{-}_{+} &= -\frac{S_{13}}{\Delta},\qquad
t^{-}_{-} = \frac{S_{11}}{\Delta}.
\end{aligned}
\end{equation}
with reflection amplitudes
\begin{equation}
\begin{aligned}
r^{+}_{+} &= \frac{S_{21}S_{33}-S_{23}S_{31}}{\Delta},\\[2pt]
r^{+}_{-} &= \frac{S_{41}S_{33}-S_{43}S_{31}}{\Delta},\\[2pt]
r^{-}_{+} &= \frac{S_{11}S_{23}-S_{13}S_{21}}{\Delta},\\[2pt]
r^{-}_{-} &= \frac{S_{11}S_{43}-S_{13}S_{41}}{\Delta}.
\end{aligned}
\end{equation}\par
Since the two propagating modes have different group velocities, the transmission and reflection probabilities are determined from the current density
\begin{equation}
\mathbf{J}=v_f\,\Psi^{\dagger}\boldsymbol{\sigma}\Psi.
\end{equation}\par
Accordingly, the transmission and reflection probabilities for the propagating modes are
\begin{equation}
T^{s}_{s'}=\frac{|\mathbf{J}^{\,s'}_{\mathrm{tra}}|}{|\mathbf{J}^{\,s}_{\mathrm{in}}|},
\qquad
R^{s}_{s'}=\frac{|\mathbf{J}^{\,s'}_{\mathrm{ref}}|}{|\mathbf{J}^{\,s}_{\mathrm{in}}|}.
\end{equation}
To ensure probability conservation, the coefficients satisfy
\begin{equation}
\sum_{s'} \left( T^{s}_{s'} + R^{s}_{s'} \right) = 1.
\end{equation}
For example, for an incident wave in the $k^{+}$ mode,
\begin{equation}
T^{+}_{+} + T^{+}_{-} + R^{+}_{+} + R^{+}_{-} = 1.    
\end{equation}

\end{document}